\documentclass[sigconf]{acmart}

\usepackage[english]{babel}
\usepackage[utf8]{inputenc}
\usepackage{blindtext}
\usepackage{subfigure}

\renewcommand\footnotetextcopyrightpermission[1]{} 
\setcopyright{none}

\acmDOI{}

\acmISBN{}

\acmConference[Submitted for review to LANCOMM STUDENT WORKSHOP]{}
\acmYear{2019}
\copyrightyear{}

\acmPrice{}

\begin{document}
\title{\texttt{RDNA Balance}: Load Balancing by Isolation of Elephant Flows using Strict Source Routing}


\author{Rodolfo V. Valentim, Rodolfo S. Villaca, Moises R. N. Ribeiro, Magnos Martinello}
\email{rodolfo.valentim@aluno.ufes.br}
\affiliation{
    \institution{Federal University of Esp\'{\i}rito Santo (UFES)}
    \state{Esp\'{\i}rito Santo}
    \country{Brazil}
}

\author{Cristina K. Dominicini, Diego R. Mafioletti}
\email{cristina.dominicini@ifes.edu.br}
\affiliation{%
    \institution{Federal Institute of Education, Science \\and Technology of Esp\'{\i}rito Santo (IFES)}
    \state{Esp\'{\i}rito Santo}
    \country{Brazil}
}

\renewcommand{\shortauthors}{Valentim et al.}

\begin{abstract}
    Data center networks need load balancing mechanisms to dynamically serve a large number of flows with different service requirements. However, traditional load-balancing approaches do not allow the full utilization of network resources in a simple, programmable, and scalable way. In this context, this paper proposes \texttt{RDNA Balance} that exploits elephant flow isolation and source routing in core nodes. Flow classification operations are performed on the edge using features of the OpenFlow protocol. The results show that with this approach it is possible to provide a simple, scalable, and programmable load balancing for data centers.
\end{abstract}

\maketitle

\vspace{-0.05cm}
\section{Introduction}
\label{sec-introduction}

Data Center (DC) administrators have to attend to several flows with conflicting network requirements due to the great variety of services and applications that needs to be executed in DCs \cite{Kandula2009}. These problems demand the existence of an efficient mechanism for Traffic Engineering (TE). Such TE mechanism needs to be able to classify the existing flows, isolate flows with conflicting network requirements and assign paths in a way that meets the flow requirements without compromising other flows.

Usually, Data Center Networks (DCNs) have complex and rigid routing mechanisms relying in a table-based routing. This approach has known scalability issues \cite{Jin:2016:YDC:2890955.2890967} and brings a high level of complexity to manage the network state. 

Modern DCNs present intense communication between servers and load balancing mechanisms and have to deal with routing in the core, which can be quite complicated. On average, every 1 $ms$, 100 new flows arrive the DCN; flows up to 25$s$ are responsible for more than half the amount of traffic; and only 0.1\% of the total of flows lasts longer than 100$s$, accounting for almost 20\% of the data volume transferred \cite{Kandula2009}. Another important observation is that a small fraction of the links experiences much greater losses than the rest of the network. It is thus demonstrated that alternative paths can be used to avoid losses and to improve the quality of service provided for network flows in a DC \cite{Benson}.

Source Routing (SR) mechanisms decrease the size of routing tables and reduce the overload in the control plane when compared with traditional approaches \cite{martinello:2014}. Strict Source Routing (SSR) methods allows to specify the entire underlay path of a flow using a routing information in the packet. According to \cite{Jyothi2015}, such routing solutions using SSR reduce the amount of flow rules installed in the network core. The lower the number of flows the better is the scalability.

The use of SSR in a DCN is a promising way to tackle TE problems. The Residue Defined Network Architecture (RDNA) is a network architecture proposed by \cite{Liberato2018} that explores the Residue Number System (RNS) and SSR to perform packet routing in the network core. RDNA separates core and edge elements of the network. The core elements are simple and route packages using module operations without tables.

\section{\texttt{RDNA Balance}}
\label{sec-rdn-balance}

This work elaborates a solution for the load balancing problem in DCNs, named \texttt{RDNA Balance}. We leverage the SSR mechanism used in the RDNA architecture, based on fabric core nodes and programmable edge nodes, to perform load balancing at the source in a reactive and centralized way. The mechanism aims to isolate elephant flows from mice flows to improve the bandwidth available to the elephant flows while decreasing the latency experienced by mice flows.

\begin{figure*}[!t]
\subfigure[RDNA Balance architecture]{\label{fig:arquiteturaRDNABalance}\includegraphics[height=0.28\textwidth]{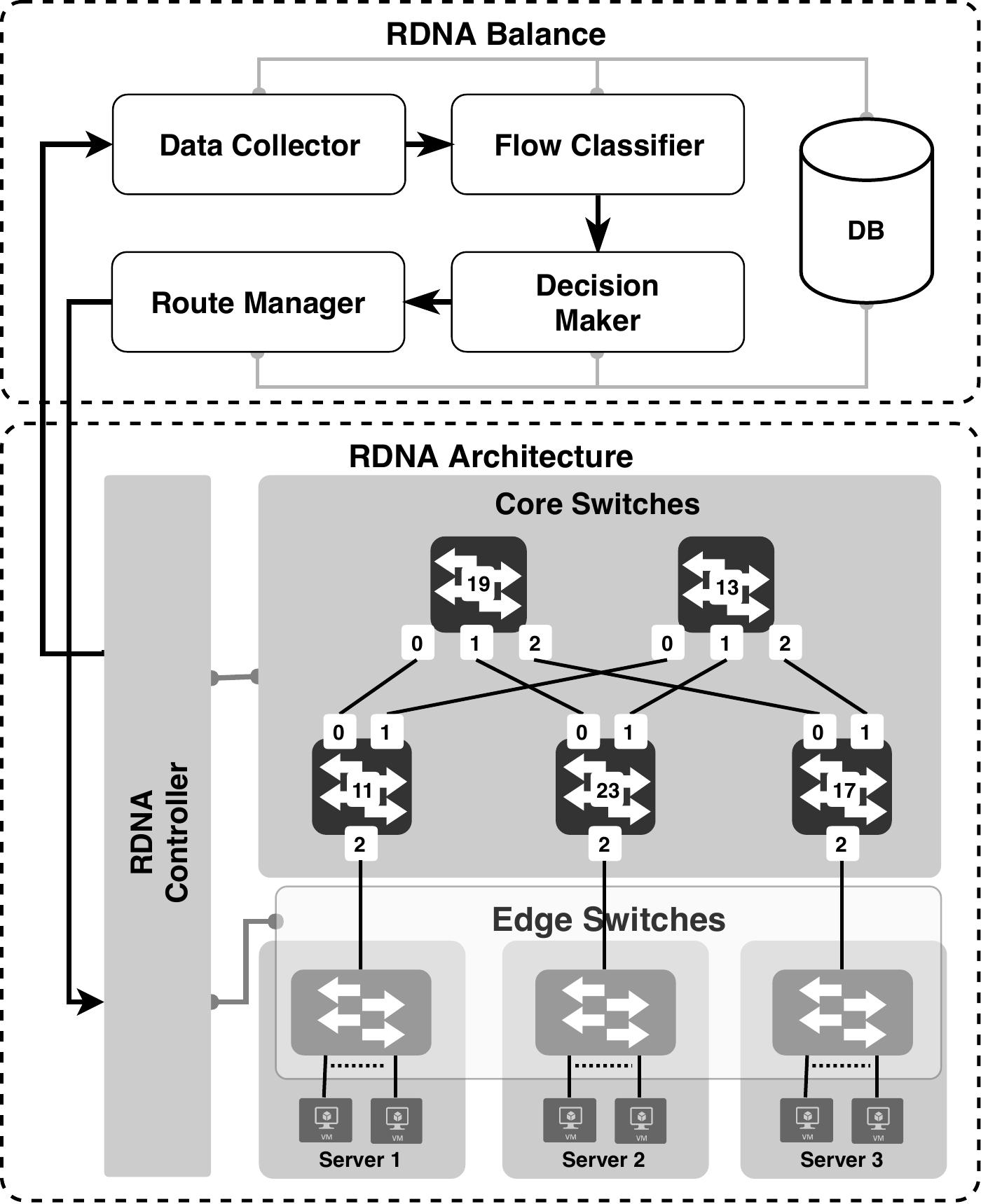}} \hspace{2mm}
\subfigure[Package loss at migration]{\label{fig:graficoCostOfMigration}\includegraphics[height=0.28\textwidth]{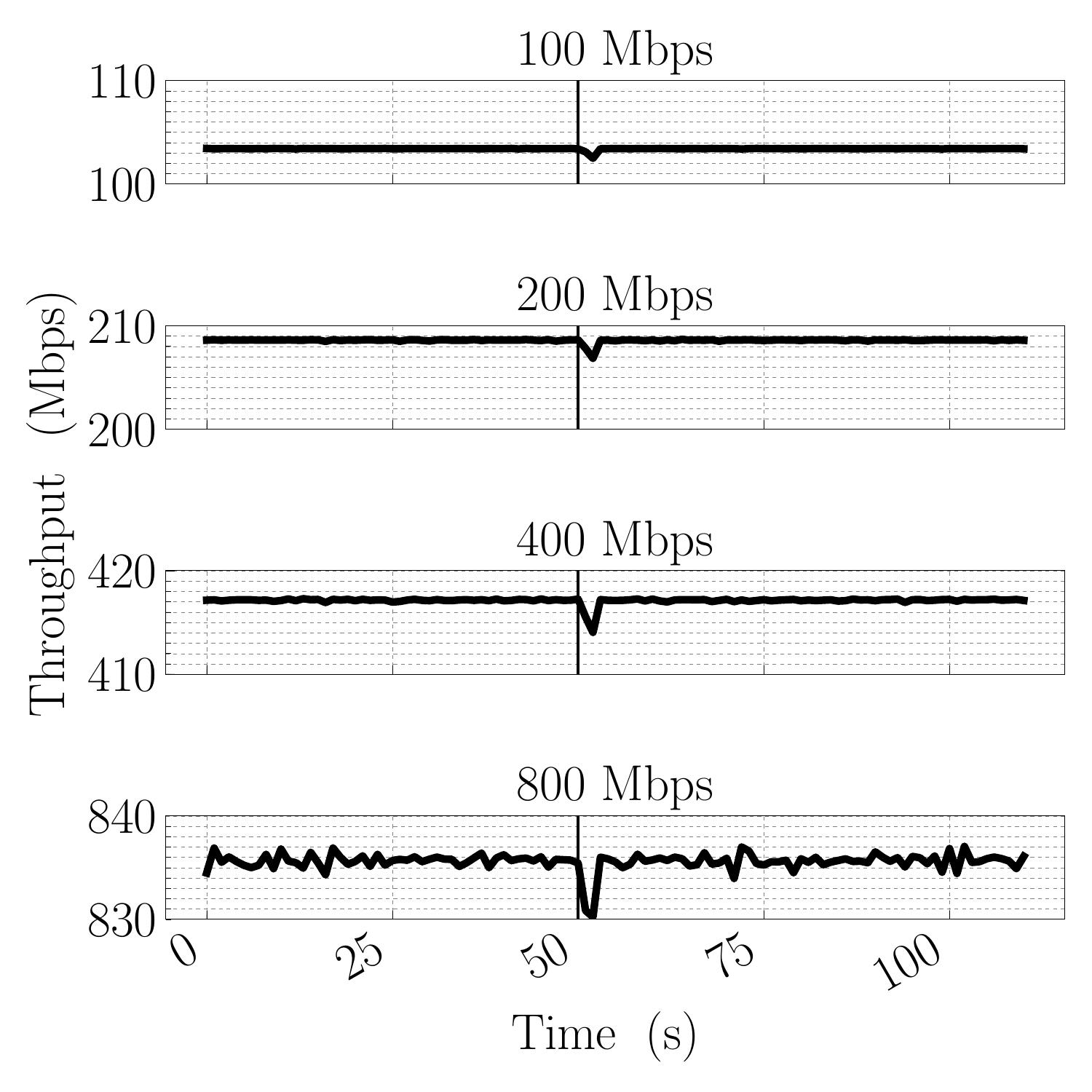}}
\subfigure[Test scenario]{\label{fig:test_scenario}\includegraphics[height=0.28\textwidth]{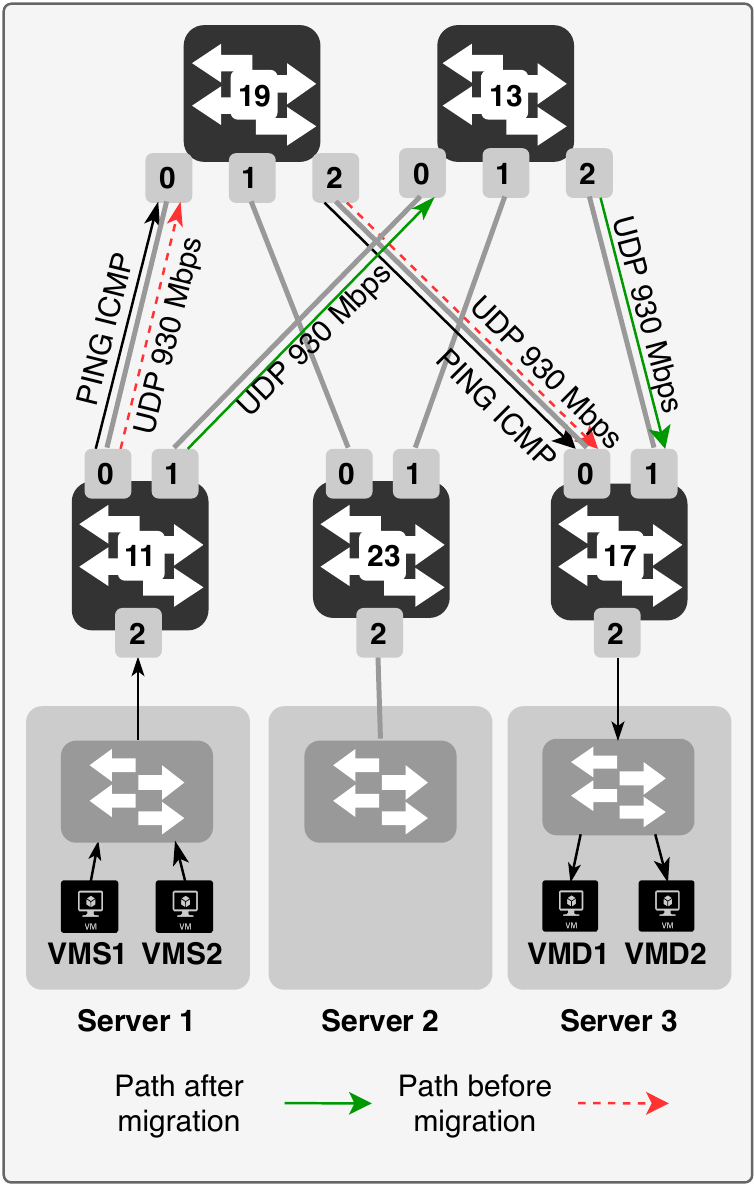}} \hspace{2mm}
\subfigure[Mice flow latency]{\label{fig:latencia}\includegraphics[height=0.28\textwidth]{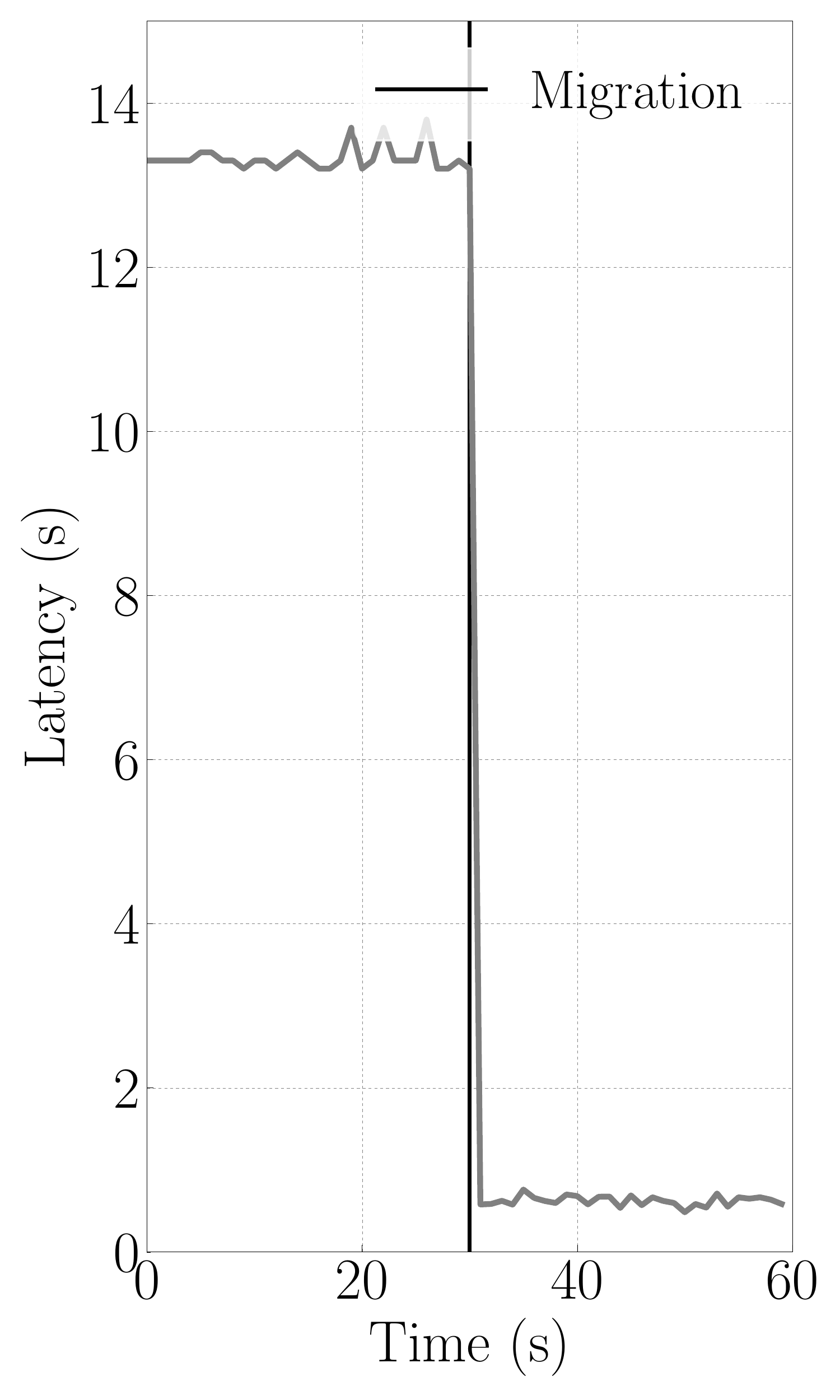}} \hspace{2mm}
  \vspace{-0.3cm}
  \caption{RDNA Balances test and results}
  \vspace{-0.2cm}
  \label{fig:sfcscenario}
\end{figure*}

The RDNA architecture is composed by the three elements represented in the Figure \ref{fig:arquiteturaRDNABalance}: i) RDNA Controller, a logically centralized controller used to configure polices and manage the switches; ii) edge switches, which insert route identifiers in the packets specifying a path to the flow; and iii) core switches that forward packets based on a \textit{modulo} operation (remainder of division) between the route identifier and the switch identifier. For instance, a package with route identifier $R = 133$ when enters a switch with identifier 11 will be forwarded according to the \textit{modulo} operation $<133>_{11}$. As the result is $1$ the package will leave the Switch $11$ by the port $1$ and enter the Switch $19$. 

In Figure \ref{fig:arquiteturaRDNABalance}, the interaction between \texttt{RDNA Balance} and a RDNA based DCN is presented. In order to simplify the communication, all the interaction is done via the RDNA controller. The four functional blocks and the database represent the main functionalities required to perform an effective load balancing via \texttt{RDNA Balance}.

All functional blocks have well defined functions in the load balancing task: i) \textbf{Data Collector} monitors the DCN and collects topological and network usage data; ii) \textbf{Flow Classifier} classify flows by their  types; iii) \textbf{Decision Maker} take load balancing decisions according to the current network state and; iii) \textbf{Route Manager} acts over the network via RDNA Controller based on the load balancing decision. This work focus on the the Route Manager, i.e, the mechanisms on the network devices in order to obtain a reactive, centralized and host-based load balancing.

In \textit{RDNA Balance}, when the Decision Maker decides to migrate a flow based on the current state of the network it communicates the Route Manager. The task to migrate a flow from one route to another involves the installation of two OpenFlow rules in the edge switch source and destination. The rule in the source inserts the route ID field "Source MAC Address" in the package's Ethernet header and the rule in the destination restores the package header. In a traditional network, the task to reroute a flow would require to change in the state of all the switches in the path between source and destination which takes longer than the RDNA Balance approach and it becomes harder to manage as the number of hops in the path increases. 

A prototype was implemented using \textit{OpenStack} as the cloud manager framework. Edge and core switches were virtualized using a customized version of \textit{OpenvSwitch (OvS)}, in which the forwarding is based on \textit{modulo} operations, implemented at kernel-level by modifying the \textit{OvS} implementation. Each switch is represented as different \textit{OvS} bridges. 

The first test verified the packet loss during route migration. In this experiment, a UDP elephant stream with different bandwidth requirements was created, ranging from 100 $Mbps$ to 800 $Mbps$, doubling the throughput at each iteration. 
The original route is: $VMS1 \rightarrow S_{11} \rightarrow S_{19} \rightarrow S_{17} \rightarrow VMD2$. After 50$s$, the route is migrated to a path with the same length: $VMS1 \rightarrow S_{11} \rightarrow S_{13} \rightarrow S_{17} \rightarrow VMD2$. In Figure \ref{fig:graficoCostOfMigration} shows the throughput at the destination $VMD1$ for each one of the evaluated throughput sent by $VMS1$.

Consider the scenario proposed in Fig. \ref{fig:test_scenario} where two flows share the same link: Flow 1 uses UDP protocol with a throughput of $930 Mbps$ and is characterized as a elephant flow; Flow 2 sends a ICMP message every $1 s$ and is characterized as a mice flow. Flow 1 is generated using \textit{pktgen} which is capable of generating packets of specific sizes at a constant rate. \textit{Pktgen} generated UDP packages at a rate of $81274 pps$ (packets per second) and packet size of $1518 Bytes$. At this rate, Flow 1 saturated the physical limit of 930 $Mbps$ link. The rate and packet size where chosen according to \cite{popoviciu2008ipv6, bolla2006rfc}. 

At $30 s$, the path of Flow 1 is migrated from $VMS1 \rightarrow S_{11} \rightarrow S_{19} \rightarrow S_{17} \rightarrow VMD2$ to $VMS1 \rightarrow S_{11} \rightarrow S_{13} \rightarrow S_{17} \rightarrow VMD2$. After the migration, Flow 1 and Flow 2 do not share the same link anymore and the latency immediately decreases from 13$ms$ to 0.7$ ms$. The latency of Flow 2 during the experiment can be checked at Figure \ref{fig:latencia}. 

\vspace{-0.4cm}
\section{Conclusion}
\label{sec-conclusion}

The results show that the mechanism proposed by \texttt{RDNA Balance} is able to migrate routes with low data loss rate, without compromising the communication between servers. Besides, the results show the mechanism offers flexibility for path selection, since the migration is simple and manageable. Future work can fill the gaps in the congestion detection, flow characterization, and queue overflow detection.

\bibliographystyle{ACM-Reference-Format}
\bibliography{reference}


\begin{thebibliography}{8}


\ifx \showCODEN    \undefined \def \showCODEN     #1{\unskip}     \fi
\ifx \showDOI      \undefined \def \showDOI       #1{#1}\fi
\ifx \showISBNx    \undefined \def \showISBNx     #1{\unskip}     \fi
\ifx \showISBNxiii \undefined \def \showISBNxiii  #1{\unskip}     \fi
\ifx \showISSN     \undefined \def \showISSN      #1{\unskip}     \fi
\ifx \showLCCN     \undefined \def \showLCCN      #1{\unskip}     \fi
\ifx \shownote     \undefined \def \shownote      #1{#1}          \fi
\ifx \showarticletitle \undefined \def \showarticletitle #1{#1}   \fi
\ifx \showURL      \undefined \def \showURL       {\relax}        \fi
\providecommand\bibfield[2]{#2}
\providecommand\bibinfo[2]{#2}
\providecommand\natexlab[1]{#1}
\providecommand\showeprint[2][]{arXiv:#2}

\bibitem[\protect\citeauthoryear{Benson, Anand, Akella, and Zhang}{Benson
  et~al\mbox{.}}{2010}]%
        {Benson}
\bibfield{author}{\bibinfo{person}{Theophilus Benson}, \bibinfo{person}{Ashok
  Anand}, \bibinfo{person}{Aditya Akella}, {and} \bibinfo{person}{Ming Zhang}.}
  \bibinfo{year}{2010}\natexlab{}.
\newblock \showarticletitle{Understanding Data Center Traffic Characteristics}.
\newblock \bibinfo{journal}{{\em SIGCOMM Comput. Commun. Rev.\/}}
  \bibinfo{volume}{40}, \bibinfo{number}{1} (\bibinfo{date}{Jan.}
  \bibinfo{year}{2010}), \bibinfo{pages}{92--99}.
\newblock
\showISSN{0146-4833}
\showDOI{%
\url{https://doi.org/10.1145/1672308.1672325}}


\bibitem[\protect\citeauthoryear{Bolla and Bruschi}{Bolla and Bruschi}{2006}]%
        {bolla2006rfc}
\bibfield{author}{\bibinfo{person}{Raffaele Bolla} {and}
  \bibinfo{person}{Roberto Bruschi}.} \bibinfo{year}{2006}\natexlab{}.
\newblock \showarticletitle{RFC 2544 performance evaluation and internal
  measurements for a Linux based open router}. In \bibinfo{booktitle}{{\em High
  Performance Switching and Routing, 2006 Workshop on}}. IEEE,
  \bibinfo{pages}{6--pp}.
\newblock


\bibitem[\protect\citeauthoryear{Jin, Farrington, and Rexford}{Jin
  et~al\mbox{.}}{2016}]%
        {Jin:2016:YDC:2890955.2890967}
\bibfield{author}{\bibinfo{person}{Xin Jin}, \bibinfo{person}{Nathan
  Farrington}, {and} \bibinfo{person}{Jennifer Rexford}.}
  \bibinfo{year}{2016}\natexlab{}.
\newblock \showarticletitle{Your Data Center Switch is Trying Too Hard}. In
  \bibinfo{booktitle}{{\em Proceedings of the Symposium on SDN Research}} {\em
  (\bibinfo{series}{SOSR '16})}. \bibinfo{publisher}{ACM},
  \bibinfo{address}{New York, NY, USA}, Article \bibinfo{articleno}{12},
  \bibinfo{numpages}{6}~pages.
\newblock
\showISBNx{978-1-4503-4211-7}
\showDOI{%
\url{https://doi.org/10.1145/2890955.2890967}}


\bibitem[\protect\citeauthoryear{Jyothi, Dong, and Godfrey}{Jyothi
  et~al\mbox{.}}{2015}]%
        {Jyothi2015}
\bibfield{author}{\bibinfo{person}{Sangeetha~Abdu Jyothi}, \bibinfo{person}{Mo
  Dong}, {and} \bibinfo{person}{P~Brighten Godfrey}.}
  \bibinfo{year}{2015}\natexlab{}.
\newblock \showarticletitle{{Towards a Flexible Data Center Fabric with Source
  Routing}}.
\newblock  (\bibinfo{year}{2015}).
\newblock
\showISBNx{9781450334518}


\bibitem[\protect\citeauthoryear{Kandula, Sengupta, Greenberg, Patel, and
  Chaiken}{Kandula et~al\mbox{.}}{2009}]%
        {Kandula2009}
\bibfield{author}{\bibinfo{person}{Srikanth Kandula}, \bibinfo{person}{Sudipta
  Sengupta}, \bibinfo{person}{Albert Greenberg}, \bibinfo{person}{Parveen
  Patel}, {and} \bibinfo{person}{Ronnie Chaiken}.}
  \bibinfo{year}{2009}\natexlab{}.
\newblock \showarticletitle{{The nature of data center traffic}}. In
  \bibinfo{booktitle}{{\em Proceedings of the 9th ACM SIGCOMM conference on
  Internet measurement conference - IMC '09}}. \bibinfo{publisher}{ACM Press},
  \bibinfo{address}{New York, New York, USA}, \bibinfo{pages}{202}.
\newblock
\showISBNx{9781605587714}
\showDOI{%
\url{https://doi.org/10.1145/1644893.1644918}}


\bibitem[\protect\citeauthoryear{Liberato, Martinello, Gomes, Beldachi, Salas,
  Villaca, Ribeiro, Kanellos, Nejabati, Gorodnik, and Simeonidou}{Liberato
  et~al\mbox{.}}{2018}]%
        {Liberato2018}
\bibfield{author}{\bibinfo{person}{Alextian Liberato}, \bibinfo{person}{Magnos
  Martinello}, \bibinfo{person}{Roberta~L. Gomes}, \bibinfo{person}{Arash
  Beldachi}, \bibinfo{person}{Emilio Salas}, \bibinfo{person}{Rodolfo Villaca},
  \bibinfo{person}{Moises R.~N. Ribeiro}, \bibinfo{person}{George Kanellos},
  \bibinfo{person}{Reza Nejabati}, \bibinfo{person}{Alexander Gorodnik}, {and}
  \bibinfo{person}{Dimitra Simeonidou}.} \bibinfo{year}{2018}\natexlab{}.
\newblock \showarticletitle{{RDNA: Residue-Defined Networking Architecture
  Enabling Ultra-Reliable Low-Latency Datacenters}}.
\newblock \bibinfo{journal}{{\em IEEE Transactions on Network and Service
  Management\/}} \bibinfo{volume}{4537}, \bibinfo{number}{c}
  (\bibinfo{year}{2018}), \bibinfo{pages}{1--1}.
\newblock
\showISSN{1932-4537}
\showDOI{%
\url{https://doi.org/10.1109/TNSM.2018.2876845}}


\bibitem[\protect\citeauthoryear{Martinello, Ribeiro, Emerick Z.~de Oliveira,
  and Vitoi}{Martinello et~al\mbox{.}}{2014}]%
        {martinello:2014}
\bibfield{author}{\bibinfo{person}{Magnos Martinello}, \bibinfo{person}{M.R.N.
  Ribeiro}, \bibinfo{person}{Rafael Emerick Z.~de Oliveira}, {and}
  \bibinfo{person}{Romulo Vitoi}.} \bibinfo{year}{2014}\natexlab{}.
\newblock \showarticletitle{KeyFlow: A Prototype for Evolving SDN Toward Core
  Network Fabrics}.
\newblock \bibinfo{journal}{{\em Network, IEEE\/}}  \bibinfo{volume}{28}
  (\bibinfo{date}{03} \bibinfo{year}{2014}), \bibinfo{pages}{12--19}.
\newblock
\showDOI{%
\url{https://doi.org/10.1109/MNET.2014.6786608}}


\bibitem[\protect\citeauthoryear{Popoviciu, Hamza, Van~de Velde, and
  Dugatkin}{Popoviciu et~al\mbox{.}}{2008}]%
        {popoviciu2008ipv6}
\bibfield{author}{\bibinfo{person}{Chip Popoviciu}, \bibinfo{person}{A Hamza},
  \bibinfo{person}{G Van~de Velde}, {and} \bibinfo{person}{D Dugatkin}.}
  \bibinfo{year}{2008}\natexlab{}.
\newblock \bibinfo{booktitle}{{\em IPv6 benchmarking methodology for network
  interconnect devices}}.
\newblock \bibinfo{type}{{T}echnical {R}eport}.
\newblock


\end{thebibliography}

\end{document}